\documentclass[conference]{IEEEtran}
\ifCLASSINFOpdf
\else
\fi
\usepackage{booktabs}
\usepackage{amsmath,amssymb,amsfonts}
\usepackage{algorithmic}
\usepackage{graphicx}
\usepackage{textcomp}
\usepackage{xcolor}
\usepackage{hyperref}
\usepackage{array}
\usepackage{siunitx}
\usepackage{booktabs}
\usepackage{float}
\usepackage{hyperref}
\usepackage{listings} 
\usepackage{color} 
\usepackage{xcolor}
\usepackage{hyperref}
\definecolor{codegreen}{rgb}{0,0.6,0}
\definecolor{codegray}{rgb}{0.5,0.5,0.5}
\definecolor{codepurple}{rgb}{0.58,0,0.82}
\definecolor{backcolour}{rgb}{0.95,0.95,0.92}
\definecolor{codeblue}{rgb}{0.25, 0.5, 0.75}

\usepackage{listings}
\usepackage{xcolor}
\usepackage{array}

\lstdefinestyle{mystyle2}{
    backgroundcolor=\color{backcolour},   
    commentstyle=\color{green},
    keywordstyle=\color{blue},
    stringstyle=\color{purple},
    basicstyle=\ttfamily\footnotesize, 
    breakatwhitespace=false,         
    breaklines=true,                 
    captionpos=b,                    
    keepspaces=true,                 
    numbers=none,                                  
    showspaces=false,                
    showstringspaces=false,
    showtabs=false,                  
    tabsize=1
}

\lstdefinestyle{mystyle}{
    backgroundcolor=\color{backcolour},   
    commentstyle=\color{codeblue},
    keywordstyle=\color{magenta},
    stringstyle=\color{codepurple},
    basicstyle=\ttfamily\footnotesize,
    breakatwhitespace=false,         
    breaklines=true,                 
    captionpos=b,                    
    keepspaces=true,                 
    numbers= none,                           
    showspaces=false,                
    showstringspaces=false,
    showtabs=false,                  
    tabsize=1
}

\begin{document}
%
\title{Better Python Programming for all: With the focus on
Maintainability}

\author{
    \IEEEauthorblockN{Karthik Shivashankar}
    \IEEEauthorblockA{University of Oslo\\
    karths@ifi.uio.no}
    \and
        \IEEEauthorblockN{Antonio Martini}
    \IEEEauthorblockA{University of Oslo\\
    antonima@ifi.uio.no}
}


%


\maketitle

\begin{abstract}
This study aims to enhance the maintainability of code generated by Large Language Models (LLMs), with a focus on the Python programming language. As the use of LLMs for coding assistance grows, so do concerns about the maintainability of the code they produce. Previous research has mainly concentrated on the functional accuracy and testing success of generated code, overlooking aspects of maintainability.

Our approach involves the use of a specifically designed dataset for training and evaluating the model, ensuring a thorough assessment of code maintainability. At the heart of our work is the fine-tuning of an LLM for code refactoring, aimed at enhancing code readability, reducing complexity, and improving overall maintainability.

After fine-tuning an LLM to prioritize code maintainability, our evaluations indicate that this model significantly improves code maintainability standards, suggesting a promising direction for the future of AI-assisted software development.
\end{abstract}


%
\IEEEpeerreviewmaketitle

\section{Introduction}

 The advent of Code Large Language Models (LLMs) started a new era in automated programming, offering unparalleled assistance in generating syntactically correct and functionally robust code. Despite their increasing adoption as coding assistants  \cite{Liu2023}, concerns regarding the maintainability of the code produced by these models persist \cite{hou2023large}.  Sustainable software development hinges on quality aspects like maintainability, which is crucial for the long-term success of software projects, affecting factors such as technical debt and the cost of future modifications\cite{software-ac-uk}. While existing research has extensively explored LLM-generated code's functional accuracy and testing efficacy, maintainability has often been overlooked \cite{xu2022systematic}. This gap highlights a pressing need to re-evaluate and enhance how code is generated, emphasising its long-term sustainability and adherence to best coding practices \cite{hou2023large}. 

 The significance of code quality becomes evident when considering the concept of technical debt, which draws an analogy to financial debt. It characterises the eventual expenses incurred when opting for code that is easy to implement in the short term rather than choosing the best overall solution \cite{lenarduzzi2021}. Within the realm of Code LLMs, creating code lacking maintainability or conformity to established standards contributes to the accumulation of technical debt \cite{ALSOLAI2020106214}. This, in turn, results in higher long-term costs associated with maintenance and scalability \cite{fan2023large}.
 
 In response to this challenge, our study seeks to explore the potential of fine-tuning Code LLMs to prioritise code maintainability, specifically within the context of Python programming. By focusing on attributes such as code readability, complexity reduction, and overall maintainability, we aim to build an AI-assisted software development towards generating code that is not only functional but also easy to understand, modify, and extend.

\subsubsection{Research Questions}

Central to our investigation is the inquiry into the extent to which fine-tuning LLMs can enhance their capacity to assess and improve the maintainability of Python code.

RQ1: "How does fine-tuning augment LLMs' capability to generate Python code with higher maintainability, and can the improvements be measured?" 

 This question underpins our broader objective of setting new benchmarks for Code LLM development, ensuring that the output meets functional requirements and adheres to high standards of code quality and maintainability.

\subsubsection{Motivation}

The motivation for exploring RQ1  stems from a recognised gap in the current literature concerning the maintainability of LLM-generated code  \cite{xu2022systematic}\cite{hou2023large}. High-quality, maintainable code significantly facilitates updates, debugging, and scalability, thus minimising long-term maintenance costs and fostering ongoing innovation \cite{siy2001modern, 7816501}.  As Code LLMs continue to gain traction for their efficiency in streamlining coding tasks, it becomes imperative to ensure that their outputs are operationally effective and maintainable in the long run.  This study addresses this critical need by fine-tuning LLMs with a specific emphasis on maintainability and conducting rigorous performance evaluations.

\subsubsection{Our contributions are twofold:
}

\begin{enumerate}
    \item Development of a Custom Dataset: Addressing the gap identified in RQ1, we've curated a dataset focusing on the maintainability of Python code, emphasising readability, complexity and modifiability or ease of refactoring the code. This dataset is engineered for developing and fine-tuning LLMs, ensuring that generated code aligns with maintainability best practices. By making this dataset publicly available, we facilitate broader research and development efforts to embed maintainability principles directly within Code LLM outputs.

    \item Empirical Evaluation of Fine-tuned Model: We introduce an experimental method leveraging our extended custom dataset to evaluate fine-tuned  LLMs on maintainability. This method extends conventional functional testing by scrutinising the maintainability of the code generated, thereby offering a comprehensive assessment of the model's output. This approach lays the groundwork for future evaluations of Code LLMs, promoting a balanced consideration of functionality and code maintainability.
\end{enumerate}

Our fine-tuned maintainability model shifts focus from general practice in AI-assisted code, which may require significant adjustments post-code generation, to a proactive approach where code quality and maintainability are prioritised early in the development cycle. This strategy helps reduce the need for extensive maintenance and technical debt, making it crucial to integrate Code LLMs into professional settings where functionality and maintainability are essential.

\section{Background}

\subsection{Maintainability}

Maintainability is a crucial quality attribute determining the ease of understanding, modifying, and extending software \cite{software-ac-uk}. It is crucial for controlling costs and ensuring the adaptability of software systems over time. Misra et al have highlighted the importance of evaluating how design and coding practices influence software maintainability, which can lead to significant reductions in maintenance costs and improvements in quality  \cite{misra2005modeling}. Adherence to maintainable coding standards promotes code readability and consistency, which are fundamental for collaborative development efforts and the long-term viability of software projects \cite{siy2001modern, 7816501}. However, there is a gap in research regarding the compliance of Code LLMs with such standards, calling into question their practicality in actual software development scenarios. Addressing this gap is essential for integrating Code LLMs into the industry, ensuring that the code they generate is functional, maintainable, and aligned with professional coding practices. Our research contributes to understanding and improving code maintainability through LLMs, but it does not encompass all dimensions or metrics associated with maintainability. This acknowledgement underscores the complexity of code maintainability and the ongoing need for research in this area.

\subsection{Parameter Efficient fine tuning (PEFT)}

Parameter-efficient fine-tuning (PEFT) has become increasingly relevant in machine learning, particularly for adapting large pre-trained models to specific tasks with minimal computational overhead \cite{peft_huggingface}. PEFT methods have been developed to achieve strong task performance while updating a significantly smaller number of parameters compared to full model fine-tuning. This cost-effective approach addresses the challenge of making informed design choices on the PEFT configurations, such as architecture and the number of tunable parameters \cite{peft_huggingface} \cite{zhou2023autopeft}.

Furthermore, as foundation models scale, the efficiency of adaptation methods like PEFT becomes critical. Pu et al. provided a comprehensive benchmark of various PEFT techniques, offering a framework for selecting the optimal fine-tuning techniques based on task type and data availability. They also challenged the notion that PEFT techniques always converge faster than full tuning, especially in low data scenarios, and optimised these techniques for better performance with fewer parameters \cite{pu2023empirical}. PEFT with QLoRA is an efficient finetuning method for large language models significantly reducing memory usage. It allows finetuning a 65B parameter model on a single GPU using innovations like 4-bit models, double quantisation, and paged optimisers \cite{qlora}.
The Hugging Face  Supervised Fine-tuning Trainer (SFT Trainer) API allows for fine-tuning language models. It supports features like training for text completion tasks, formatting input prompts, packing datasets for efficiency, and controlling pre-trained model parameters. Advanced usage includes training adapters using the PEFT library, allowing users to train only parts of a model and share these smaller, adapted models\cite{huggingface_sft}. We use SFT trainer and PEFT with QLoRA to fine-tune the open-source Code LLM model, which we discuss later in the Methodology section. 

\section{Related Works}

Large Language Models (LLMs) for code continue to grow, with limited studies focusing on code maintainability. Building upon the notion of code quality, Zhuo (\cite{zhuo2023large}) introduces a framework that assesses the quality of LLM-generated code. This framework works by  ensuring the code aligns with human standards, 
In the context of code testing, Xiong, Guo, and Chen (\cite{xiong2023program}) explore how LLMs can assist in program synthesis by generating test cases. This ability to produce not only functional code but also comprehensive tests underscores the potential of LLMs to work on other qualities or attributes.
Shirafuji et al. (\cite{shirafuji2023exploring}) investigate the robustness of LLMs in programming problem-solving and reveal the sensitivity of these models to prompt engineering. This insight is integral to standardisation, suggesting that with suitable prompts, LLMs could be steered to produce code that adheres to specific standards, hence promoting consistency and standardisation in coding practices.
Lastly, Le et al. (\cite{le2022coderl}) proposed "CodeRL," integrating a pre-trained model with deep reinforcement learning for program synthesis. While their focus is on functional correctness, the principles underlying CodeRL could be adapted to emphasise the generation of standard-compliant code.
Linking these contributions together, it becomes evident that the field of LLMs for code is expanding beyond mere code generation, delving into broader aspects of code quality. However, despite these advances, there appears to be a gap in research explicitly targeting the maintainability of code generated by LLM. To our knowledge, no studies have yet explored this area in depth. Our research fills this gap, offering unique insights and methodologies for enhancing code maintainability, particularly in Python, using LLMs.
\section{Methodology}
\begin{figure}[h]
    \centering
    \includegraphics[width=1\linewidth]{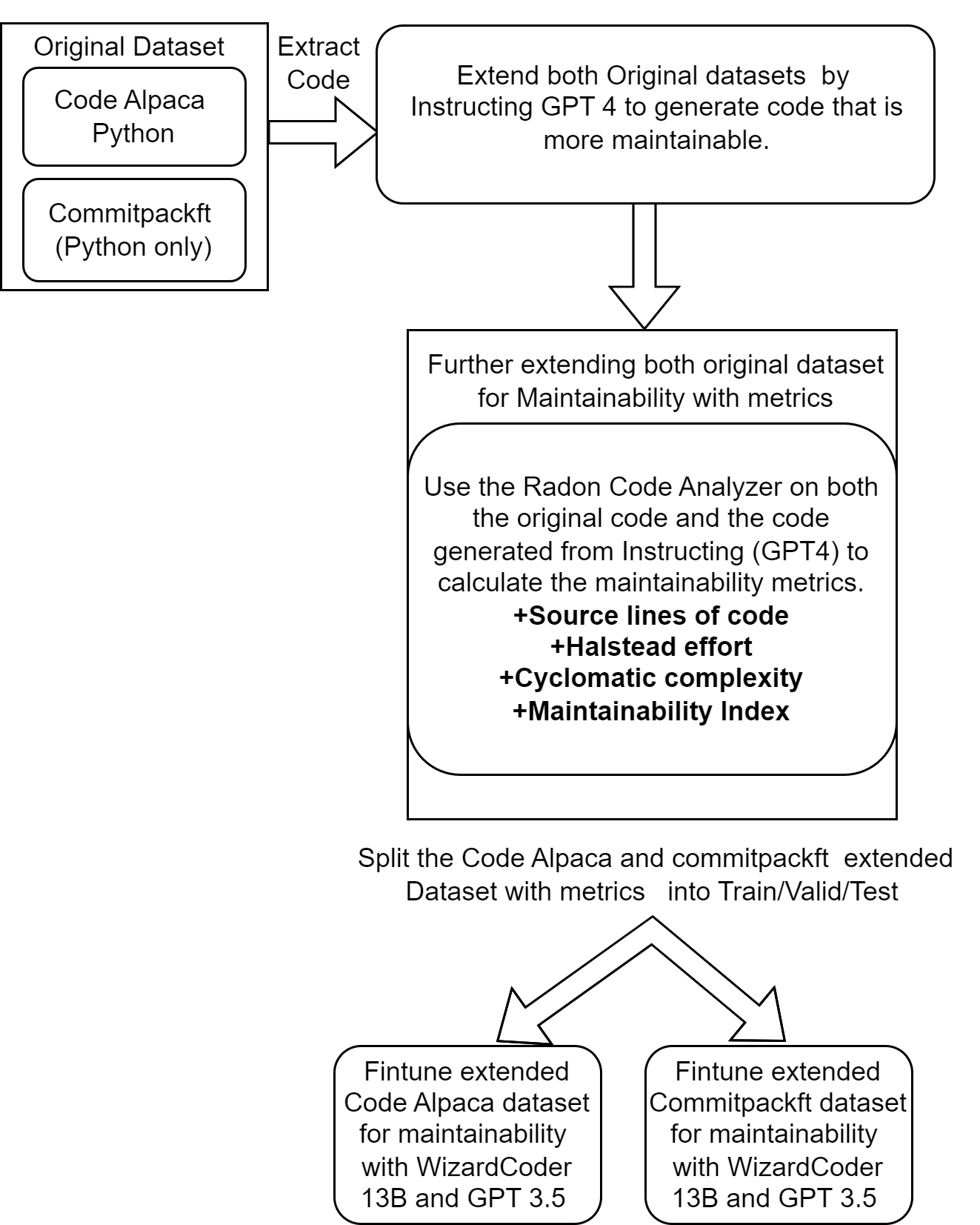}
    \caption{Steps used  for Curating Datasets and  Fine-tuning LLM }
    \label{fig:methodology}
\end{figure}

This research employs a systematic methodology to investigate LLMs' generation and evaluation of maintainable Python code. The approach is structured into several vital steps (Figure \ref{fig:methodology}), each addressing the main RQ1. The methodology combines dataset preparation, model fine-tuning as shown in Figure \ref{fig:methodology}, and comprehensive evaluation techniques to assess the capability of LLMs in producing code that functions correctly and adheres to high maintainability standards.
We use a mixed-method approach that combines practical experimentation with qualitative analysis. This choice is driven by the need to evaluate the multifaceted nature of code maintainability, which encompasses code's syntactic and functional correctness and aspects such as readability, complexity, and adherence to best practices. The methodology is designed to be replicable and rigorous, allowing for a clear assessment of the impact of fine-tuning LLMs on code maintainability. 

        \textbf{Comparative Analysis and Evaluation}
The comparative analysis aims to quantify the improvements in code maintainability due to fine-tuning the LLM. This analysis involves comparing the maintainability metrics (SLOC, Halstead effort, Cyclomatic complexity, and Maintainability Index) of the original code in the dataset and to that code generated by the base models versus the fine-tuned models. The metrics are calculated for each code sample in the test split, providing a basis for statistical analysis to identify significant trends and improvements.
The evaluation process, a crucial part of our work, is designed to validate the functional correctness and maintainability of the generated code. This involves using CodeBertScore \cite{zhou2023codebertscore} to compare the similarity and functional correctness of the code generated by the fine-tuned models against the original code samples. Additionally, the utility of the fine-tuned models is assessed through invaluable feedback from expert Python programmers, who play a pivotal role in evaluating the generated code's quality and adherence to best practices like code readability, modularity, and reusability, which are essential for reducing maintenance costs. 
\subsection{Selecting Dataset for Maintainability }
For finetuning Code LLMs, the initial step typically involves using various original code samples as the primary input. These code samples are from the CommitPackFT dataset (Python subset) \cite{muennighoff2023octopack}and CodeAlpaca-18k (Python) dataset  \cite{python_code_instructions_18k_alpaca}. This immerses the models in various authentic coding situations, with the imperfections and inconsistencies they would encounter in the practical coding environment.

\paragraph{Diversity of data and rationale for choosing the Dataset:}

CommitPackFT(Python subset) \cite{muennighoff2023octopack} offers many real-world programming scenarios and high-quality instructional commit messages. Its critical feature is its focus on high-quality commit messages (that serve as instruction) paired with code. This unique feature makes it particularly suitable for tasks like file-level and repository-level refactoring, which is pivotal for understanding code changes and intentions in software development.

CodeAlpaca (Python) \cite{python_code_instructions_18k_alpaca} provides a diverse set of coding examples and instructional data essential for developing models that can follow instructions in code generation tasks.
This dataset stands out due to its variety in code types, including snippets and functions, and its accompanying instructions and inputs. This diversity provides a rich set of coding examples, making it a valuable resource for instruction-following models. The CodeAlpaca dataset was inspired by a project aiming to build an instruction-following LLaMA model for code generation based on the Stanford Alpaca model \cite{wang-etal-2023-self-instruct}. 

\begin{table}[htbp]
\centering
\caption{Dataset split}
\label{tab:dataset_division}
\begin{tabular}{lccc}
\hline
Dataset & Training Set & Validation Set & Testing Set \\ 
\hline
CommitpackFT& 13,983 (78\%) & 1,998 (11\%) & 2,000 (11\%) \\
CodeAlpaca& 12,927 (80\%) & 1,617 (10\%) & 1,614 (10\%) \\
\hline
\end{tabular}
\end{table}
Table \ref{tab:dataset_division} outlines the distribution of two datasets, CommitpackFT (Python Subset) and CodeAlpaca Python, into training, validation, and testing sets, which are crucial for fine-tuning and evaluating the model. These datasets are mainly used to fine-tune models like GPT-3.5 and WizardCoder with 78\% and 80\% of the data, respectively, while both validation and testing sets each comprise about 10-11\%, essential for model evaluation and assessing code maintainability improvements.

\subsection{Instructing GPT4 to generate Maintainable code}
To enhance the code maintainability of the code samples from the chosen dataset, we employ the GPT4 model, distinguished for its prowess in code generation among various tasks \cite{openai2023gpt4}. We instruct or prompt  GPT-4 with detailed context, including original code alongside maintainability metrics, aiming to generate refactored code that preserves the original functionality while optimising maintainability, as shown in Listing 1. Instruction tuning, a method in natural language processing, diverges from traditional supervised training by integrating instructions into input-output pairs, thereby augmenting the model's adaptability across tasks. This technique has shown promising outcomes, notably in models trained on datasets like Stanford Alpaca  \cite{alpaca_crfm_2023}.  We adapted this Alpaca style to our Maintainability context. Given the variety of prompting strategies and their sensitivity to model parameters, we offer our prompt as a recipe for adaptation and leave further exploration of prompting techniques and strategies to the research community.

\begin{minipage}{0.48\textwidth}
\begin{lstlisting}[language=Python, caption=Maintainability Prompt]
Maintainbility_prompt = f"""You are a Python expert specialising in code optimisation.
    Your main task is to refactor the given Python code to improve upon the listed metrics:
    Source Lines of Code (SLOC), Effort, and Maintainability Index (MI), while retaining the original functionality
### Input:
{original_code}
### Context:
Objective
- Improve Source Lines of Code (SLOC): Lower numbers are generally better without compromising readability or functionality.
- Improve Maintainability Index (MI): Higher scores are desired.
- Reduce Effort: Lower numbers are preferred.
Original Metrics
- Source Lines of Code (SLOC): {original_SLOC}
- Maintainability Index (MI): {original_maintainability_index}
- Effort: {original_effort}
Provide only the refactored version of the code with comments on what changes are made on the code and do not provide the metrics.
### Response:
{refactored_code_optimised_maintainability}*
"""\end{lstlisting} \label{maintain_prompt}
\end{minipage}
 
\paragraph{Rationale for using the GPT4 model to generate a Training dataset for maintainability?}
LLMs, including GPT4, are pivotal in generating high-quality, maintainable code at scale. Instruction Tuning (IT) has gained traction in training LLMs on specialised datasets, comprising instructions and outputs, to refine the models' output alignment with user instructions \cite{NEURIPS2023_camels,zhang2023instruction}. Research highlights the effectiveness of both curated and synthetic data from LLMs, evidenced by projects like Stanford Alpaca \cite{wang-etal-2023-self-instruct} and Evol Instruct \cite{xu2023wizardlm}, and the utility of synthetic data in instruction tuning across LLMs is underscored in literature \cite{zhang2023instruction}. This approach is becoming increasingly popular in academic and industrial settings to boost LLM performance in various contexts \cite{wang-etal-2023-self-instruct, Rosenbaum2022_CLASP, Rosenbaum2022_LINGUIST}. Our project leverages this innovative approach, focusing on fine-tuning an LLM for code maintainability through a bespoke custom extended dataset. This strategy mirrors the growing adoption of synthetic data for instruction tuning in LLMs, showcasing its expansive potential in boosting LLM performance, especially in enhancing code maintainability.
In Listing 1, you can see the maintainability prompt we used to instruct GPT4 to curate the custom dataset and for instruction fine-tuning LLM  to generate maintainable Python code.

\subsection{Evaluating the functional similarity of the generated GPT 4 code to the original code}

\begin{table}[h!]
\centering
\caption{CodeBERTScore for Validation split}
\label{table:codebert_validation}
\begin{tabular}{@{}l@{\hspace{3pt}}c@{\hspace{3pt}}c@{\hspace{3pt}}c@{\hspace{3pt}}c@{}}
\toprule
Dataset & P(mean \scriptsize{(std)}& R mean \scriptsize{(std)}& F1 mean \scriptsize{(std)}& F3 mean \scriptsize{(std)}\\
\midrule
CodeAlpaca& 0.915 \scriptsize{(0.068)} & 0.908 \scriptsize{(0.065)} & 0.910 \scriptsize{(0.061)} & 0.908 \scriptsize{(0.063)} \\
Commitpackft& 0.938 \scriptsize{(0.068)} & 0.937 \scriptsize{(0.068)} & 0.937 \scriptsize{(0.066)} & 0.937 \scriptsize{(0.067)} \\
\bottomrule
\end{tabular}
\end{table}

In Table \ref{table:codebert_validation}, we present an evaluation of generated code (from Instructing GPT 4)  to that of the reference code ( Original code) from two datasets, namely CodeAlpaca and Commitpackft, using the CodeBERTScore metrics \cite{zhou2023codebertscore}. The evaluation metric from the CodeBERTScore metric comprises precision (P), recall (R), F1, and F3 scores. CodeBERTScore utilizes the semantic understanding capabilities of pre-trained models, particularly CodeBERT \cite{feng-etal-2020-codebert}, to compute soft similarities between tokens in the generated and reference code, considering the programmatic context. This enables a nuanced evaluation by acknowledging semantic roles and functional similarity, even amidst syntactic differences between code snippets. These metrics range from 0 to 1, where 0 indicates no alignment, and 1 indicates perfect alignment between the generated and reference codes. Table \ref{table:codebert_validation} reveals high levels of functional correctness and semantic similarity between generated and reference codes, with scores around 0.91 to 0.94 on a scale from 0 to 1. 

Precision (P) assesses the accuracy of the functionalities implemented in the generated code, with a higher score indicating a greater proportion of the functionalities being correctly implemented. For example, a precision score of 0.938 in the Commitpackft dataset means that, on average, 93.8\% of the functionalities implemented in the generated code accurately reflect those in the reference code. Recall (R) measures how comprehensively the generated code captures the necessary functionalities present in the reference code. A recall score of 0.937 suggests that, on average, 93.7\% of all functionalities in the reference code are also implemented in the generated code, indicating a high level of completeness.F1 score is the harmonic mean of precision and recall, providing a single metric that balances both the accuracy and completeness of the generated code.F3 score gives more weight to recall, emphasizing the importance of covering all necessary functionalities in the generated code. The standard deviations associated with these scores, such as (0.068), indicate the variability of the scores across different instances in the dataset. A lower standard deviation means that the scores are more consistent across instances.
This indicates that the generated code is functionally and semantically similar to the original code. This signifies that, on average, the code generation process is able to produce code that is not only syntactically correct but also effectively mirrors the intended functionality and completeness of the reference implementations.

\subsection{Enhancing the Dataset with a Maintainability Metric}

For each piece of code in the original dataset, we compute maintainability metrics—namely, Maintainability Index (MI), Cyclomatic Complexity (CC), Halstead Effort (HE), and Source Lines of Code (SLOC) (refer to Table \ref{table:software-metrics}). These metrics are calculated for both the original code samples and those generated by instructing GPT-4, with the resultant maintainability metrics subsequently incorporated into the new fields within the dataset. This augmentation process is meticulously applied to all code samples, ensuring the dataset is comprehensively enhanced with these critical metrics for further evaluation and testing. To obtain these metrics, we utilised the Radon Python package, which provides an extensive suite for assessing the complexity and maintainability of Python code \cite{radon}. We selected Radon because of its widespread use in industry tools like Codacy, Code Climate, and CodeFactor. 
 \begin{table}[h!]
  \caption{Software Metric for Maintainability \cite{radon}}
  \label{table:software-metrics}
  \centering
  \begin{tabular}{m{2.0cm} m{5.5cm}}
    \toprule
    \textbf{Metric} & \textbf{Description} \\
    \midrule
    Maintainability Index (MI) & Measures the ease of maintaining the code. \\
    Cyclomatic Complexity (CC) & Indicates the number of linearly independent paths through a program. \\
    Halstead Effort (HE) & Reflects the effort required to comprehend the code. \\
    Source Lines of Code (SLOC) & Counts the number of lines in the code. \\
    \bottomrule
  \end{tabular}
\end{table}

\subsubsection{Rationale for Metric Selection:}

The utilisation of Radon and the metrics listed in Table \ref{table:software-metrics} is based on their proven ability to measure key aspects of software maintainability effectively. These metrics are recognised for their utility in assessing the quality and maintainability of code, both in industry and academic settings \cite{mikejo5000_2023, radon}. The MI is particularly valuable for estimating the effort needed for software maintenance and upgrades, which is crucial for forecasting maintenance costs. CC and HE offer detailed insights into the code's complexity and the mental effort required for its understanding. The inclusion of SLOC highlights its significance in evaluating the size and maintainability challenges of software projects \cite{sjoberg2012questioning}.

These metrics are preferred in the evaluation of software maintainability because they provide objective, quantifiable measures that allow for the comparison of different aspects of software quality, such as complexity, maintenance effort, and manageability \cite{mikejo5000_2023,radon}. 
All the metrics chosen are recognised for their effectiveness in assessing code maintainability, Additionally, we incorporated the Maintainability Index (MI), a composite metric widely used in industry tools such as Microsoft Visual Studio. MI combines lines of code, cyclomatic complexity, and Halstead measures to provide a comprehensive view of code maintainability. 

Conversely, metrics like cohesion and coupling , which evaluate the internal consistency of software modules and their interdependencies, are essential for designing maintainable software. These metrics were not used as they are more applicable at the project level with multiple files and complex architectures, whereas our focus was on individual files and  their qualitative nature renders them subjective and less precisely measurable than quantitative metrics \cite{woodward1993difficulties} \cite{cohesion}. The focus on MI, CC, HE, and SLOC underscores the practicality of using quantifiable metrics for tangible evaluation and enhancement of software maintainability. Nonetheless, it is recognised that these metrics may not fully encompass all facets of software maintainability.

\begin{itemize}
    \item The Maintainability Index ( MI) combines several metrics into a single score. High maintainability means less time and resources are needed for future modifications or debugging. It is also important to note that a Higher MI score equals higher maintainability. 
Radon's implementation of the Maintainability Index is still a very experimental metric. Radon derivative \cite{radon}:
\\

{\small
  $ MI = \max\left[0, 100 \left(\frac{171 - 5.2 \ln(V) - 0.23 G - 16.2 \ln(L) + 50 \sin(\sqrt{2.4 C})}{171}\right)\right] $
\\

where \( V \) is Halstead Volume, \( G \) is Cyclomatic Complexity, \( L \) is Source Lines of Code (SLOC), and \( C \) is the percentage of comment lines \cite{radon}.
}

    \item CC is the McCabe metric that counts the number of independent paths through the code.

    \item Halstead's effort and other Halstead metrics use operator and operand counts to measure various aspects of code complexity.

Halstead's effort \(E\), the computational cost of software development, is defined as:
{\small
\[ E = D \times V \]
with Difficulty \(D\):
\[ D = \frac{\eta_1}{2} \times \frac{N_2}{\eta_2} \]
And Volume \(V\):
\[ V = N \times \log_2 \eta \]
Where \(\eta_1\) and \(\eta_2\) are the counts of distinct operators and operands, \(N_1\) and \(N_2\) are the total counts of operators and operands, and \(N\) and \(\eta\) are the sum of \(N_1\) and \(N_2\), and \(\eta_1\) and \(\eta_2\) respectively \cite{radon}.
}
    \item SLOC measures the number of lines of code that omit comments.
\end{itemize}

\begin{table}[h]
\centering
\caption{Comparison of Metrics on the training split data}
\label{tab:Training}
\begin{tabular}{lcccc}
\toprule
& \multicolumn{2}{c}{GPT4( instruct)} & \multicolumn{2}{c}{Original code} \\
\cmidrule(lr){2-3} \cmidrule(lr){4-5}
Metrics & Mean & Median & Mean & Median \\
\midrule
SLOC & 18.55 & 18.0 & 21.15 & 21.00 \\
HE & 27.19 & 1.0 & 36.96 & 2.38 \\
MI Score & 92.87 & 100.0 & 85.56 & 93.73 \\
CC Score & 4.45 & 3.0 & 4.32 & 3.00 \\
\bottomrule
\end{tabular}
\end{table}

Table \ref{tab:Training} compares various maintainability metrics in the training split between code generated by instructing GPT-4 with improved maintainability characteristics (Using Maintainability Prompt from Listing 1) and original code from two Python datasets, 'Commitpackft' and 'CodeAlpaca' respectively, offering a detailed view of GPT-4's impact on code quality and complexity.
\begin{itemize}
    \item SLOC: The GPT-4 generated code has fewer lines on average (18.55 mean) than the original code (21.15 mean), indicating more concise code generation.
    \item HE: The effort required for GPT-4's code is lower (mean of 27.19) than that for the original code (mean of 36.96), suggesting GPT-4's efficiency in creating less complex solutions.
    \item MI Score: GPT-4 outperforms the original code with a higher mean MI score (92.87 vs. 85.56), signifying better maintainability.
    \item CC Score: The CC scores are comparable between GPT-4 and the original code, with a slight increase for GPT-4 (mean of 4.45 vs 4.32), indicating a negligible difference in code complexity.
\end{itemize}
These metrics collectively demonstrate GPT-4's capability to generate code that is not only more maintainable but also requires less effort to understand and modify while maintaining complexity levels comparable to human-written code. This provides a comprehensive view of the qualitative differences in code generated by GPT-4 compared to the original dataset. Such analysis is crucial in understanding the practical utility of AI in programming, going beyond syntactic correctness to evaluate the real-world applicability of AI-generated code in software development and maintenance.

\subsection{Fine-tuning for Maintainability  }

In this paper, fine-tuning LLMs for generating Python code focuses on improving the maintainability of the generated Python code. To achieve this, we use two models: We chose WizardCoder 13B (WizardCoder-Python-13B-V1.0), which is based on the LLama family of models  \cite{code_llama_meta_2023} because it performs very well across many coding benchmarks like HumanEval and fits our GPU constraints and budget \cite{HumanEval} \cite{luo2023wizardcoder}  
The other model we chose was OpenAI's GPT-3.5(gpt-3.5-turbo-1106) \cite{OpenAI2023ChatGPT} model, which is a closed-source model. Consequently, we do not have much information on the model. We chose GPT 3.5 because of its similar performance to WizardCoder in HumanEval \cite{HumanEval}. OpenAI provides an API to fine-tune the GPT 3.5 model. This choice allows for a comprehensive comparison between open-source and closed-source models, offering insights into the strengths and limitations of each in coding tasks, thus enriching the study's depth.

\subsection{Fine-tuning techniques employed for Maintainability}

\subsubsection{WizardCoder 13B}

The fine-tuning process for WizardCoder 13B \cite{luo2023wizardcoder} incorporates a sophisticated approach that leverages the Supervised Fine-tuning Trainer (SFT) from Hugging Face \cite{huggingface_sft}, coupled with the PEFT library \cite{feng2023peftser}. The SFT technique is particularly effective in directing the model towards following instructions and aligning the model to generate code that adheres to high standards of maintainability for our case. The PEFT helps adapt pre-trained language models to specific applications with minimal fine-tuning. It fine-tunes only a few extra parameters, reducing computational and storage demands. This approach is cost-effective compared to the full fine-tuning of large-scale models. Recent PEFT techniques deliver performance similar to complete fine-tuning methods \cite{peft_huggingface}. 
Additionally, incorporating the QLoRA \cite{qlora} technique alongside PEFT  further reduced the memory requirement to run these models. This is achieved using quantization techniques. SFT with PEFT  and QLoRA allows us to fine-tune  WizardCoder 13B on an A100 GPU (40 GB). Using SFT, the model is exposed to examples from our extended datasets, which include the input original code, additional contexts,  instruction and the desired output ( Refactored code ) optimised for maintainability. The model's parameters are adjusted to minimize the difference between its outputs and the correct answers \cite{huggingface_sft}. This is achieved using the maintainability prompt from Listing 1  alongside our extended dataset.

\subsubsection{GPT-3.5}
In contrast, GPT-3.5 is fine-tuned using OpenAI's API, providing a more simplistic but effective approach. The API allows the model to be trained on a range of examples that represent good coding practices where, again, the model is exposed to include both the input code, additional contexts, instruction and the desired outputs, which are refactored for maintainability. Using the maintainability prompt from Listing 1 alongside our extended custom dataset. 
Open-source models like WizardCoder allow extensive modification and adaptation, though they require more maintenance effort from users. OpenAI's GPT-3.5 API \cite{OpenAI2023ChatGPT}, while not open source, provides a simple and easy-to-use high-level API to send fine-tuning jobs. This makes it versatile and user-friendly, but it can be costlier compared to other solutions. The choice between closed and open source depends on the specific needs, budget, and desired control over the software for researchers and developers. 

\section{Data availability}
We have made the replication package publicly available at \href{https://zenodo.org/doi/10.5281/zenodo.10153875}{\textbf{https://zenodo.org/doi/10.5281/zenodo.10153875}} \footnote{\url{https://zenodo.org/doi/10.5281/zenodo.10153875}}. It contains all the model artefacts, including final model weights, checkpoints and the associated Code for training and evaluating the WizardCoder and GPT 3.5 model on various metrics. The evaluation script can generate all the maintainability metrics reported in this paper.  We also make all the datasets used in this paper available in this replication package. All the code is seeded to ensure reproducibility and reliability of our result. We have made our best effort to anonymise all the data and code. 

\section{Results}

\subsection{ RQ1. Fine-tuning LLM to generate Python code with higher maintainability  }

The primary aim of this research question is to critically evaluate the impact of fine-tuning LLMs on their capability to produce maintainable Python code. 
Following the fine-tuning phase, the model enters a crucial testing stage. In this stage, a portion of the dataset, which remained unused during the training phase (Test split), is presented to the model. This testing aims to gauge the model's ability to apply the coding pattern and characteristics it has learned to new and unseen examples of Python code. This evaluation is essential to determine whether the model can generalize its understanding of high code quality, focusing on maintainability. It also tests the model's capacity to handle various coding scenarios and complexities representing real-world programming environments. This stage is vital in assessing how fine-tuning LLMs can enhance their ability to generate Python code for improved maintainability.

\begin{table}[!ht]
\centering
\caption{Comparison of Original and Refactored Bubble Sort Implementations}
\begin{tabular}{| m{0.455\textwidth} |}
\hline
\textbf{Original Code from dataset} \\
\hline
\begin{lstlisting}[language=Python]
def bubbleSort(arr): 
    n = len(arr) 
    for i in range(n-1):
        for j in range(0, n-i-1):
            if arr[j] > arr[j+1]:
               arr[j],arr[j+1] = arr[j+1],arr[j]
\end{lstlisting}
\\
\hline
\textbf{Refactored Code from FT model (WizardCoder 13B)} \\
\hline
\begin{lstlisting}[language=Python]
def bubbleSort(arr): 
    n = len(arr) 
    from itertools import product
    for i, j in product(range(n-1), range(n-1)):
        if arr[j] > arr[j+1]: 
            arr[j], arr[j+1] = arr[j+1], arr[j]
# Changes made:
# 1. Imported the product function from itertools module to combine the two for loops into one.
# 2. This improves readability.
# 3. The functionality remains the same as the original code.
\end{lstlisting}
\\
\hline
\textbf{Original code metrics} \\
\hline
SLOC: 6 \\
CC Score: 4 \\
Effort: 209.28 \\
MI Score: 69.58 \\
\hline
\textbf{Refactored Code from FT model metrics} \\
\hline
SLOC: 6 \\
CC Score: 3 \\
Effort: 194.40 \\
MI Score (without comments): 70.49 \\
\hline
\end{tabular}
\label{table:bubble_sort_comparison}
\end{table}

This Table \ref{table:bubble_sort_comparison} presents a simple comparative example of original and refactored implementations of the bubble sort algorithm, post-fine-tuning (FT)  of a WizardCoder. The refactoring process demonstrates an exciting use of the "itertools.product" function, a modern approach to simplifying the nested loops in the original code. It also removes a For loop from the Original code implementation. 
The metrics provided include SLOC, CC Score, Effort (calculated using Halstead's complexity measures), and the MI Score. The unchanged SLOC suggests that compactness was maintained, while a reduction in the CC Score from 4 to 3 indicates a simplification in control flow complexity, potentially making the code easier to understand and modify. The effort metric shows a slight decrease, suggesting the refactored code requires less cognitive effort to comprehend. The MI Score increase from 69.58 to 70.49 in the refactored code signifies an improvement in maintainability. This underscores the LLM’s capability to not only learn coding patterns but also apply these insights to produce more maintainable code, illustrating the practical benefits of fine-tuning LLMs for specific programming tasks. 

The model is fine-tuned to produce maintainable Python code, complemented by "Changes Made" comments ( as shown in Table \ref{table:bubble_sort_comparison} ) detailing each modification and its impact on code quality. This approach not only enhances the maintainability of the code but also educates developers on best practices, aiding their understanding and application of writing maintainable code techniques.

\begin{figure}[h]
    \centering
    \includegraphics[width=1\linewidth]{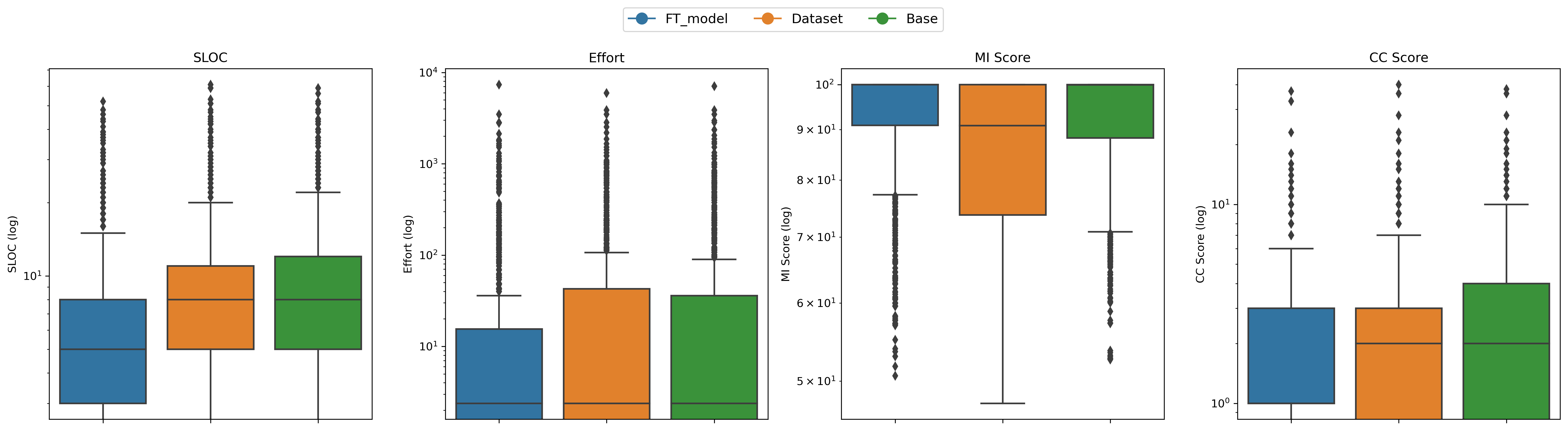}
    \caption{Comparing the distribution  of metrics for CodeAlpaca Test split  with WizardCoder 13B (Blue(FT model), Orange(Dataset) and Green(Base Model)}
    \label{fig:2}
\end{figure}

\begin{figure}[h]
    \centering
    \includegraphics[width=1\linewidth]{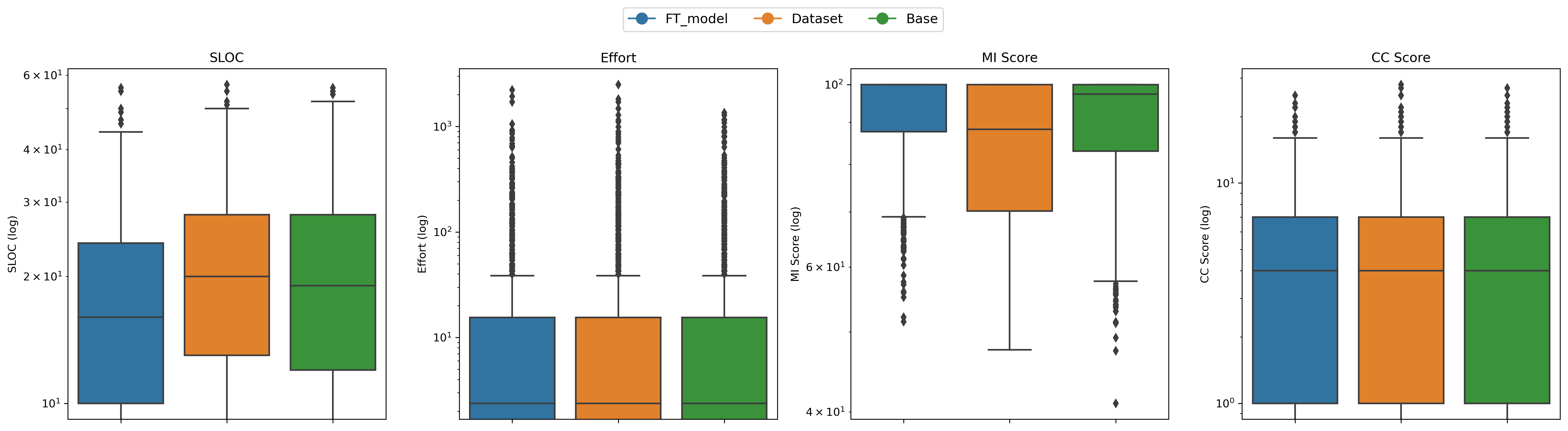}
    \caption{Comparing the distribution  of metrics for Commitpackft  Test split with WizardCoder 13B}
    \label{fig:3}
\end{figure}

\begin{figure}[h]
    \centering
    \includegraphics[width=1\linewidth]{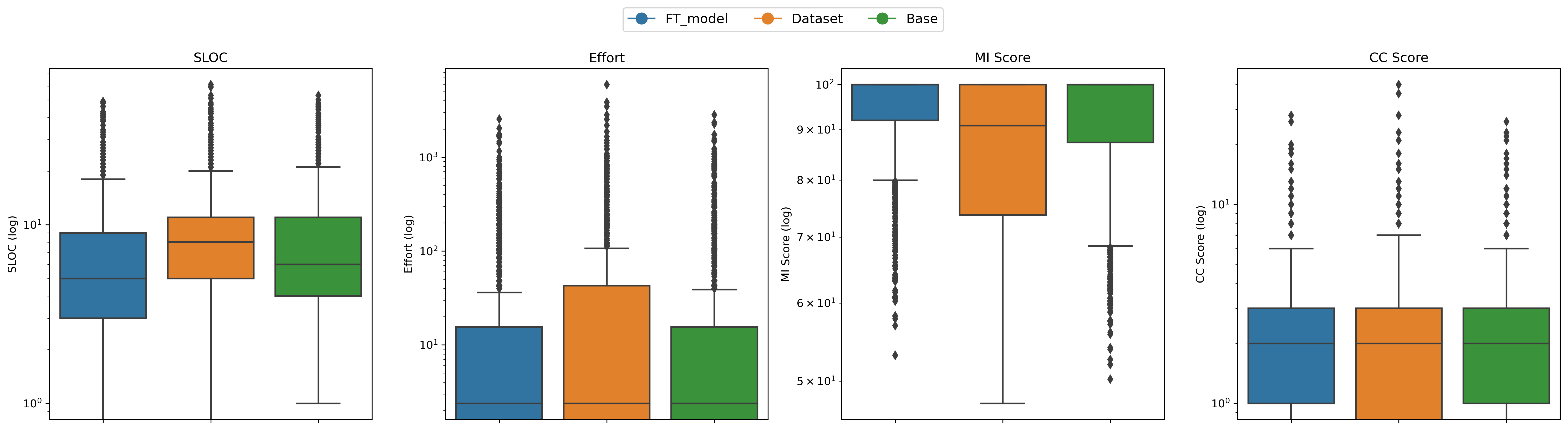}
    \caption{Comparing the distribution  of metrics for the CodeAlpaca Test split  with GPT 3.5 }
    \label{fig:4}
\end{figure}

\begin{figure}[h]
    \centering
    \includegraphics[width=1\linewidth]{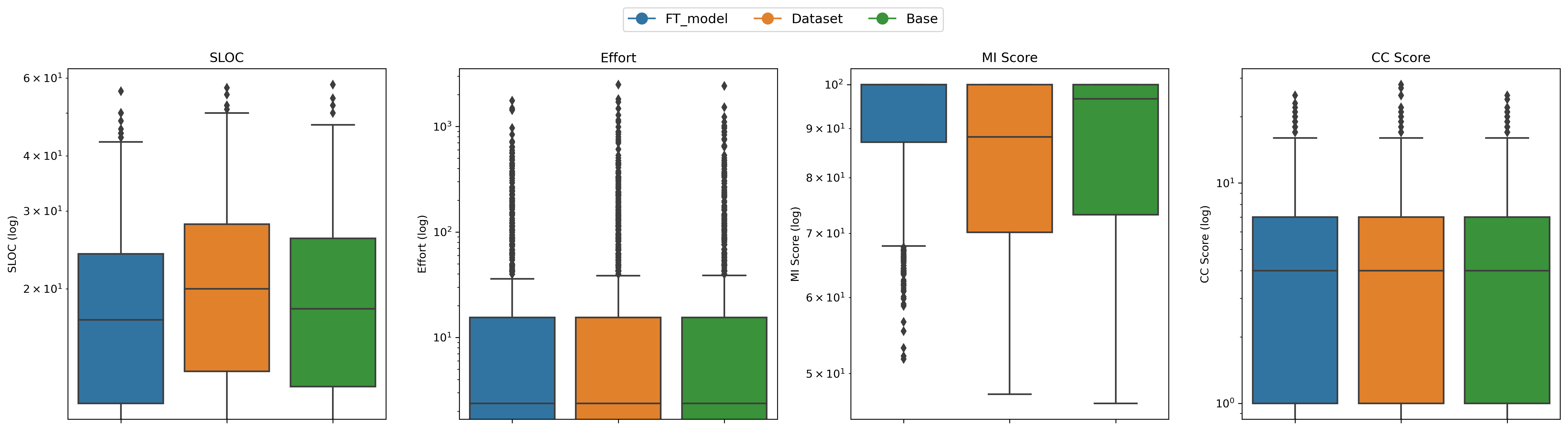}
    \caption{Comparing the distribution  of metrics for Commitpackft  Test split with GPT-3.5}
    \label{fig:5}
\end{figure}

\paragraph{Analysis of Distribution Box Plot}
The box plot in Figures \ref{fig:2}, \ref{fig:3}, \ref{fig:4}, and \ref{fig:5}  provides a comprehensive view of the performance across the Original Dataset, Base Model and Fine-Tuned (FT)  Model. The box plot visualises the distribution of maintainability metrics for the Original Dataset, Base Model, and Fine-Tuned Model. Interpretation should focus on the median (central line in each box), range (interquartile range indicated by the box), and outliers (diamonds beyond the whiskers). All the metrics reported in the Tables below in this section are mean values for the entire test split. Therefore, this box plot is crucial as it illustrates the range and distribution of the maintainability metrics, including a comparison of median values and the extent of variability and outliers within this box plot, thereby complimenting the overall comparison. This visual representation is essential for understanding the full scope of the models' effectiveness and the impact of fine-tuning on code quality.

    The Tables \ref{tab:Wizard_python} and \ref{tab:wizard_commits} focus on evaluating the effectiveness of fine-tuned WizardCoder13B (FT Model)  to Base WizardCoder13B Model in generating maintainable Python code, using maintainability metrics from the CodeAlpaca and Commitpackft Test splits. For a more detailed comparison, we have also included the Original Dataset maintainability metrics on these dataset test splits.

\begin{table}[h]
\centering
\caption{Comparing WizardCoder 13B Maintainability  Metrics from CodeAlpaca Test Split }
\label{tab:Wizard_python}
\begin{tabular}{lcccc}
\hline
Metrics & Dataset & Base Model & FT Model &  \% (\scriptsize Base vs. FT Model) \\
\hline
SLOC & 9.30 & 9.59 & 6.52 & 46.96 \\
Effort & 77.58 & 86.19 & 58.54 & 47.24 \\
MI Score & 86.46 & 92.70 & 93.79 & -1.16 \\
CC Score & 2.42 & 2.65 & 2.22 & 19.03 \\
\hline
\end{tabular}
\end{table}

Key Results from Table \ref{tab:Wizard_python}, 
In this CodeAlpaca test split, the fine-tuned WizardCoder13B model significantly decreased SLOC by 46.96\%, indicating more compact code generation. The Effort also decreased by 47.24\%, suggesting enhanced maintainability. The MI Score increased from 92.70 to 93.79 when compared to the base model. Since the MI Score ranges up to 100, with higher scores indicating better maintainability, this increase of approximately 1.16\% suggests an improvement in code maintainability. Thus, the change is positive, contrary to what might be implied by the negative sign in "-1.16\%".The CC Score improved by 19.03\%, denoting simpler and more maintainable code.

The box plot in Figures \ref{fig:2} shows the maintainability metrics where the Fine-Tuned (FT) Model generally outperforms the Base Model, as indicated by lower median values in SLOC and Effort, suggesting more efficient and less complex code. The MI Score is slightly higher for the FT Model, indicating better code quality, while the CC Score's lower median suggests simpler code. This visual comparison highlights the FT Model's enhanced code maintainability.

\begin{table}[h]
\centering
\caption{Comparing WizardCoder 13B Maintainability  Metrics from Commitpackft Test split }
\label{tab:wizard_commits}
\begin{tabular}{lcccc}
\hline
Metrics & Dataset & Base Model & FT Model & \% (\scriptsize Base vs. FT Model) \\
\hline
SLOC & 20.82 & 20.34 & 17.44 & 16.65 \\
Effort & 41.14 & 37.27 & 31.11 & 19.78 \\
MI Score & 84.94 & 90.25 & 92.57 & -2.51 \\
CC Score & 4.53 & 4.67 & 4.46 & 4.85 \\
\hline
\end{tabular}
\end{table}

Key Results from Table \ref{tab:wizard_commits}, 
In this Commitpackft test split, the fine-tuned model achieved a 16.65\% decrease in SLOC, demonstrating more concise code generation. The Effort needed was reduced by 19.78\%, indicating improved intelligible code. The MI Score marginally increased, suggesting enhanced code maintainability. The CC Score showed a minor improvement, indicating a slight decrease in code complexity.

The box plot in Figures \ref{fig:3} shows a visual comparison that highlights the FT Model's enhanced code maintainability compared to the Base Model and Dataset.

    The Tables \ref{tab:gpt_python} and  \ref{tab:gpt_commits} focus on evaluating the effectiveness of fine-tuned  GPT-3.5 (FT Model)  to Base GPT 3.5 Model in generating maintainable Python code, using maintainability metrics from the CodeAlpaca and Commitpackft Test splits. For a more detailed comparison, we have also included the Original Dataset maintainability metrics on these dataset test splits.

\begin{table}[h]
\centering
\caption{Comparing GPT3.5  Results and Metrics from CodeAlpaca Test Split for Maintainability }
\label{tab:gpt_python}
\begin{tabular}{lcccc}
\hline
Metrics & Dataset & Base Model & FT Model & \% (\scriptsize  Base vs FT Model) \\
\hline
SLOC & 9.30 & 8.51 & 6.95 & 22.41 \\
Effort & 77.58 & 59.62 & 47.83 & 24.65 \\
MI Score & 86.46 & 92.48 & 94.61 & -2.26 \\
CC Score & 2.42 & 2.45 & 2.38 & 2.80 \\
\hline
\end{tabular}
\end{table}

Key Results Table\ref{tab:gpt_python}, 
In this CodeAlpaca test split, the fine-tuned (FT)  GPT-3.5  demonstrated notable improvements in various metrics. The SLOC showed a 22.41\% decrease, indicating more concise code generation. The Effort also decreased by 24.65\%, suggesting enhanced maintainability. The MI Score saw a slight increase, indicating better code maintainability. CC Score improved by 2.80\%, denoting simpler and more maintainable code.

The box plot in Figures \ref{fig:4} shows a visual comparison highlighting the FT Model's enhanced code maintainability in all metrics compared to the Base Model and Dataset.
\begin{table}[h]
\centering
\caption{Comparing GPT 3.5 Maintainability  Metrics from Commitpackft Test split }
\label{tab:gpt_commits}
\begin{tabular}{lcccc}
\hline
Metrics & Dataset & Base Model & FT Model & \% (\scriptsize Base vs FT Model) \\
\hline
SLOC & 20.82 & 19.38 & 18.27 & 6.06 \\
Effort & 41.14 & 34.84 & 29.94 & 16.35 \\
MI Score & 84.94 & 86.89 & 92.57 & -6.14 \\
CC Score & 4.53 & 4.62 & 4.73 & -2.35 \\
\hline
\end{tabular}
\end{table}
Key Results Table  \ref{tab:gpt_commits} , 
In this Commitpackft test split,  the fine-tuned  GPT-3.5 demonstrated notable improvements in various metrics. The SLOC showed a 6.06\% decrease, indicating a more concise code generation. The Effort also saw a decrease of 16.35\%, suggesting enhanced maintainability. However, the MI Score showed a notable increase for the FT model, jumping from 86.89 in the base model to 92.57. This significant rise in the MI Score is indicative of substantially better maintainability. On the other hand, the CC Score in the FT model increased marginally to 4.73 from the base model's 4.62, a change of -2.35\%. This slight increase in complexity may suggest a marginal rise in the intricacy of the code produced by the FT model.

The box plot in Figures \ref{fig:5} shows a visual comparison that highlights the FT Model's enhanced code maintainability compared to the Base Model and Dataset.

\section{Evaluation }

\subsection{Evaluating the functional similarity on the test split}

The evaluation of fine-tuned models on Test split using CodeBERTScores shows high functional correctness and similarity between the original reference code from the dataset and the generated code from the FT models on both Commitpack and CodeAlpaca datasets.

CodeBERTScore  \cite{zhou2023codebertscore} is a metric designed to evaluate the similarity between a reference code snippet and a generated code snippet, focusing on both syntactical and functional equivalence. Unlike traditional metrics such as BLEU  \cite{papineni2002bleu}, which mainly rely on exact token matches, CodeBERTScore leverages the contextual embeddings from a model like CodeBERT \cite{feng-etal-2020-codebert}, which is trained on both programming languages and natural language. This approach allows CodeBERTScore to understand the underlying semantics of code snippets beyond mere lexical similarities. 
The key advantage of CodeBERTScore is its ability to recognize functionally equivalent code snippets that may not share a high degree of lexical similarity. For example, it can be understood that \verb|x ** 0.5| and \verb|math.sqrt(x)| performs the same operation (calculating the square root of \verb|x|) despite having different tokens. This is a significant improvement over traditional metrics, ensuring that generated code is evaluated more accurately in terms of what it does rather than just how it is written.

CodeBERTScore represents a sophisticated approach to evaluating code generation. It prioritizes the functional and semantic accuracy of the generated code over mere lexical matching. This makes it especially suitable for applications where understanding the intent and functionality of code is crucial, such as automated code review or code synthesis tasks.
The metrics Precision (P), Recall (R), F1 score, and F3 range from 0 to 1, where 1 indicates a perfect match, and 0 indicates no match at all between the generated code and the reference code. 

\begin{table}[h!]
\centering
\caption{CodeBERTScores evaluation metric with Mean and Standard Deviation on test split}
\label{table:codebert_scores}
\begin{tabular}{@{}l@{\hspace{3pt}}c@{\hspace{3pt}}c@{\hspace{3pt}}c@{\hspace{3pt}}c@{}}
\toprule
\textbf{FT Model}& \textbf{P mean \scriptsize{(std)}}& \textbf{R mean \scriptsize{(std)}}& \textbf{F1 mean \scriptsize{(std)}}& \textbf{F3 mean \scriptsize{(std)}}\\
\midrule
GPT3.5 Commitpack& 0.93 \scriptsize{(0.07)} & 0.93 \scriptsize{(0.06)} & 0.93 \scriptsize{(0.06)} & 0.93 \scriptsize{(0.06)} \\
Wizard Commitpack& 0.94 \scriptsize{(0.06)} & 0.94 \scriptsize{(0.07)} & 0.94 \scriptsize{(0.06)} & 0.94 \scriptsize{(0.07)} \\
GPT3.5 CodeAlpaca& 0.90 \scriptsize{(0.08)} & 0.90 \scriptsize{(0.08)} & 0.90 \scriptsize{(0.08)} & 0.90 \scriptsize{(0.08)} \\
Wizard CodeAlpaca& 0.92 \scriptsize{(0.08)} & 0.91 \scriptsize{(0.07)} & 0.91 \scriptsize{(0.07)} & 0.91 \scriptsize{(0.07)} \\
\bottomrule
\end{tabular}
\end{table}

Table \ref{table:codebert_scores} presents an evaluation of the functional similarity between original and generated code using the CodeBERTScores \cite{zhou2023codebertscore} metric, which encompasses precision (P), recall (R), F1, and F3 scores. These scores reflect the nuanced capability of fine-tuned models to produce code that is not only syntactically correct but also semantically aligned and functionally equivalent to the reference code. A mean score close to 1 (e.g., 0.90 or 0.94) suggests that the generated code is very similar to the reference code in terms of functionality and semantics.  The standard deviations associated with these scores, such as (0.08), indicate the variability of the scores across different instances in the dataset. A lower standard deviation means that the scores are more consistent across instances. 

This means that the generated code is highly effective and comprehensive, with most of its functionalities correctly implemented. The results demonstrate the potential of these models in applications requiring precise and functionally accurate code generation, moving beyond mere syntactic similarity to ensure semantic and functional alignment with original code snippets.

\subsection{Evaluating and assessing the usefulness and utility  }

Our evaluation method for assessing a fine-tuned AI model as a programming assistant involved a structured session where participants with varying Python expertise from both industry and academia (A total of 11 participants answered all the questions) interacted with the FT model to complete coding tasks reflective of real-world scenarios. The tasks were self-selected by participants to ensure relevance to their actual coding practices. Feedback was gathered through a questionnaire focusing on the model's usefulness and effectiveness as an AI companion, rated on a 1 to 5 scale, with 1 being "not useful at all" and 5 being "extremely useful".  

\begin{table}[htbp]
\centering
\caption{Summary Statistics of Survey Responses}
\label{tab:summary_statistics_human}
\begin{tabular}{lccc}
\hline
Statistic & Usefulness & AI Companion & Python Exp. (yrs) \\
\hline
Mean & 3.44 & 3.60 & 10.55 \\
Median & 4.00 & 4.00 & 10.00 \\
Std. Dev. & 1.13 & 1.07 & 4.87 \\
\hline
\end{tabular}
\end{table}

Table \ref{tab:summary_statistics_human} presents summary statistics for three variables from a survey: Usefulness Rating, AI Companion Rating, and Python Experience (Years). These statistics provide insights into the central tendency and variability of the responses for each variable. We had a total response from 11 users in this survey, who answered all the questions.

For  "Usefulness Rating" and "AI Companion Rating". Both ratings have a median of 4.0(Very useful), indicating that the central tendency of user opinions is positive, suggesting the model is generally found useful and performs well as an AI companion. The medians being at 4.0 also imply that at least half of the ratings are at or above this value. The assessment is based on feedback from users with varying levels of Python programming experience, which is critical for a comprehensive evaluation of the model's performance across different expertise levels. Table \ref{tab:summary_statistics_human} reveals that users with around 10 years of experience form the largest group, suggesting that their insights are particularly valuable given their considerable expertise. 

User feedback on our Fine-tuned AI coding tool reveals that it helps write cleaner, more maintainable code, particularly for experienced Python users. Novice programmers benefit from exposure to best coding practices. Yet, some users noted the AI's suggestions occasionally do not align with specific project needs and called for faster response times and better integration with development environments for enhanced customization. Gathering human feedback grounds our evaluation in practical use and provides empirical data on the tool’s effectiveness across varied user experience levels, paving the way for further detailed exploration.

\section{Discussion}

Our fine-tuned model is designed to refine existing functional code and prioritizes suggestions that adhere to established coding standards and best practices rather than generating new code. This tool has proven to be a valuable AI companion for code refactoring, as demonstrated by positive human evaluations. 

Integrating this model into continuous integration and deployment pipelines can streamline development workflows, providing immediate feedback to mitigate technical debt accumulation. Moreover, the model serves as an instructional guide for novices, promoting the development of clean and efficient code while standardizing learning through consistent feedback. 

For the research community, it offers a platform to assess the effectiveness of automated refactoring tools and explore different refactoring strategies to enhance long-term software maintenance. This exploration of human-AI collaboration in software development could significantly elevate code quality, efficiency, and cost-effectiveness in software engineering.

The discussion of whether fine-tuning is required despite the existence of capable general models such as GPT-3.5 for code generation is crucial. Our study highlights that these models may not focus on nuanced aspects of code maintainability without targeted training. We have developed a dataset centred on Python programming, incorporating best practices and examples of highly maintainable code. This dataset is essential for fine-tuning other large language models (LLMs) to improve the maintainability of the code they generate. By utilising this dataset, our fine-tuned model demonstrates improved performance in code quality and maintainability. This not only enhances the model's practicality in software development but also enables the use of smaller, more efficient models. These models, potentially less costly in terms of computational resources, can achieve or exceed the capabilities of larger models like GPT-4, and offer additional security benefits for deployment in sensitive environments.

\subsection{Treats to validity}
This study on fine-tuning LLMs for improving the maintainability of generated Python code addresses several challenges that impact its validity and reliability. A primary internal concern is whether the training dataset comprehensively represents the Python ecosystem. Limited representation could cause models to overfit specific patterns, reducing their generalization capability. The choice of hyperparameters in fine-tuning is also crucial, as inappropriate selections can affect the models' reliability and generalizability. Externally, the study's relevance may be limited outside the Python ecosystem. The complexity of industry-level code might not be fully represented, questioning the practical utility of the models in real-world scenarios. 

In terms of construct validity, the operational definitions of maintainability may not fully encapsulate these concepts as understood in professional and academic settings. The tools and metrics used to evaluate code maintainability could have inherent biases, potentially skewing the results. The study's conclusion validity relies on robust statistical analysis. Without this, there's a risk of misinterpreting data and drawing incorrect inferences about the impact of fine-tuning. 

Reliability issues arise from potential model performance inconsistencies and the evolving nature of real-world code, known as dataset shift. These factors necessitate continuous model updates to ensure the study’s relevance in the rapidly changing field of AI-driven code generation.

Our research contributes to understanding and improving code maintainability through LLMs, but it does not encompass all dimensions or metrics associated with maintainability. This acknowledgement underscores the complexity of code maintainability and the ongoing need for research in this area.

\section{Conclusion}
This study explores the effectiveness of fine-tuning Large Language Models (LLMs) for generating maintainable Python code. Utilizing our custom extended datasets and leveraging models such as WizardCoder 13B and GPT-3.5, we have achieved notable improvements in code maintainability metrics, such as Source Lines of Code, Maintainability Index, Halstead Effort, and Cyclomatic Complexity. These enhancements highlight LLMs' potential as powerful tools in automating code refactoring processes, with the promise of advancing software development practices. Central to our investigation is a specially curated dataset designed with a focus on Python programming that can be used to fine-tune other LLMs and improve the generated code quality and maintainability. By examining the strengths and weaknesses of LLMs in producing maintainable Python code, this research contributes significantly to the fields of automated code generation and software maintainability.




%
\bibliographystyle{IEEEtran}
\bibliography{sample-base}

\end{document}